# LCLS-II AND HE CRYOMODULE MICROPHONICS AT CMTF IN FERMILAB*

C. Contreras-Martinez, E. Harms, C. Cravatta, J. Holzbauer, S. Posen, L. Doolittle[1], B. Chase, J. Einstein-Curtis[2]. J. Makara, R. Wang. Fermilab, Batavia, IL, USA
[1] Lawrence National Laboratory, Berkely, CA, USA
[2] Previously at Fermilab, now at RadiaSoft, Boulder, CO, USA

*Abstract*

Microphonics causes the cavity to detune. This study discusses the microphonics of sixteen 1.3 GHz cryomodules, 14 for LCLS-II and 2 for LCLS-II HE tested at CMTF. The peak detuning, as well as the RMS detuning for each cryomodule, will be discussed. For each cryomodule, the data was taken with enough soaking time to prevent any thermalization effects which can show up in the detuning. Each data capture taken was 30 minutes or longer and sampled at 1 kHz.

## INTRODUCTION

The LCLS-II is an X-ray FEL linac that uses 9-cell 1.3 GHz SRF cavities. The collaboration between SLAC, Fermilab, Lawrence Berkeley National Lab (LBNL), and Thomas Jefferson Lab (JLab) produced cryomodules tested at Fermilab and JLab. One of the tests done at Fermilab in the cryomodule testing facility (CMTF) facility is to record the level of microphonics. The cryomodules (CM) are shipped to SLAC in California once testing is complete at Fermilab or JLAB.

The cryomodule is roughly 13 m long and contains eight 1.3 GHz SRF cavities. Liquid helium cools the cavities to 2 K. The bandwidth of the cavities is 20 Hz, and during operation, the cavities need to experience a peak detuning of 10 Hz or less. Microphonics on the cryomodules were mitigated by using passive damping techniques. The JT cryogenic valves were modified by adding wiper rings along the valve stem to dampen thermoacoustic oscillations. The valves were also reconfigured to change the helium supply line by altering where the input connects. These techniques are described in detail by Hansen et al. [1]. Changes to the cavity 1 connection and other changes are discussed at length in Refs. [3-4]. These changes result in a substantial decrease in frequency detuning of all eight cavities. The peak detuning before these changes was as high as 200 Hz, and now it can go up to 30 Hz with these changes implemented.

Since all the changes were implemented a peak detuning below 10 Hz was observed in 63 % of all cavities as shown in Fig. 1. For 63 % of all cavities the RMS detuning was below 2 Hz (see Fig. 2). Note that these results could be unique to the CMTF environment. The supply pressure at SLAC is lower, and so is the inlet temperature. Additionally, the cryogenic plant is further apart from the linac, decreasing the vibration level. Thermalization effects can affect the cavity detuning. Therefore, the data analyzed was captured one week after the cool down to allow these effects to dissipate. The data presented in figures 1 and 2 only have two cryomodules operated at 16 MV/m while the rest are at 5 MV/m. The cavities with a peak detuning greater than 10 Hz have similar vibrations,

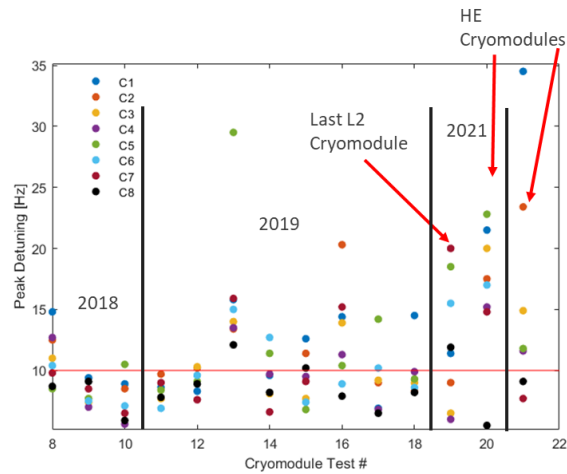

Figure 1: Cavity peak detuning of all 8 cavities from cavities tested from 2018 until 2022. These include cavities from LCLS-II and HE. The cavity peak detuning specification is 10 Hz.

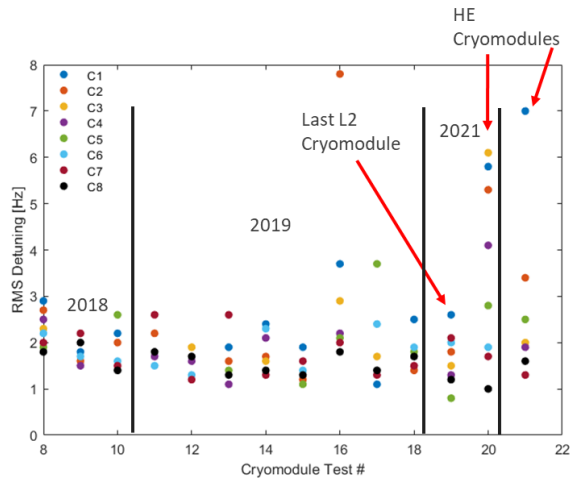

Figure 2: RMS cavity detuning from 2018 to 2022. These include cavities from LCLS-II and HE.

---

such as a 30 Hz source attributed to the Kinney vacuum pump. Other cavities are affected by cryogenics such as the 2.3 K supply pressure, harmonics, helium bath pressure, and Helmholtz mode of the helium vessel. The Helmholtz mode is affected by the liquid level inside the helium vessel; this effect is discussed in [5].

## MICROPHONICS DUE TO CRYOGENICS

### Supply Pressure Effects

All cavities tested and shown in Figure 1 and 2 have 18 Hz or 20 Hz vibrations. These vibrations were the primary contributor to the peak detuning on some cavities. A correlation between these vibrations and the 2.3 K supply line was found. This vibration was studied by changing the supply pressure of the 2.3 K line from 10 psig to 40 psig in the verification cryomodule (vCM), see Fig. 3. The results show that the 20 Hz vibration transitions to 18 Hz at 25 psig. The nominal supply pressure is 33 psig. In this data acquisition, the 18 vibration becomes stable once the nominal pressure is reached. In other cases, it is the 20 Hz vibration.

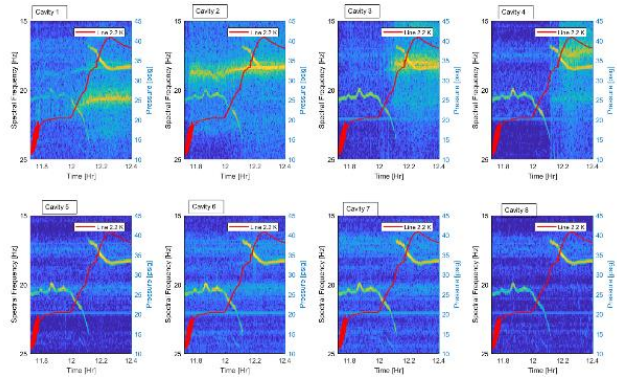

Figure 3: Spectrogram of the cavities in vCM showing the transition of 20 Hz vibration to 18 Hz by changing the supply pressure of the 2.3 K line.

### Harmonic Effects

In some of the cavities the 18 Hz vibration can lead to harmonics shown on the spectrograms in figures 4 and 5. Figure 4 shows data for CM15 which has a fundamental mode that occurs at 18 Hz with higher order modes of 36 Hz, 54 Hz, and 90 Hz. The blip at around hour 13 is due to the supply pressure for the 2.3 K line. Figure 5 shows the data for CM15, the fundamental mode is 23 Hz and the higher order mode is 46 Hz. The 23 Hz line is correlated with the 2.3 K supply pressure line. The 30 Hz line on the figure is due to the Kinney vacuum pump. The harmonics observed in the frequency detuning can be acoustic or mechanical.

The mechanical harmonics can occur from the eigenmodes of the metal structure such as the cavity string assembly, the cryomodule, and the 2 K two-phase pipe. The wavelength of the wave traveling in the metal can be estimated using the equation $\lambda = v/f$ where v is the speed of sound in the metal and f is the frequency excited. The cryomodule is made from stainless steel and the cavity is made from niobium. The speed of sound in stainless steel is 6 km/s and for niobium, it is 3.48 km/s. The fundamental frequencies of the harmonics were 18 and 23 Hz. For an 18 Hz vibration in stainless steel, a wavelength of 333 m can be excited; for niobium, a wavelength of 193 m can be excited on the metal. The 18 Hz source cannot excite harmonics on the metals since the wavelength exceeds the dimension of the entire cryomodule which is 13 m long. The speed of sound of saturated helium vapor at 2 K is 226 m/s [6], and the wavelength is 12.5 m for an 18 Hz source. In a pipe with both ends open, the length is given by $L = n\lambda/2$ where n is the harmonic order, for n=1 the length is 6.25 m.

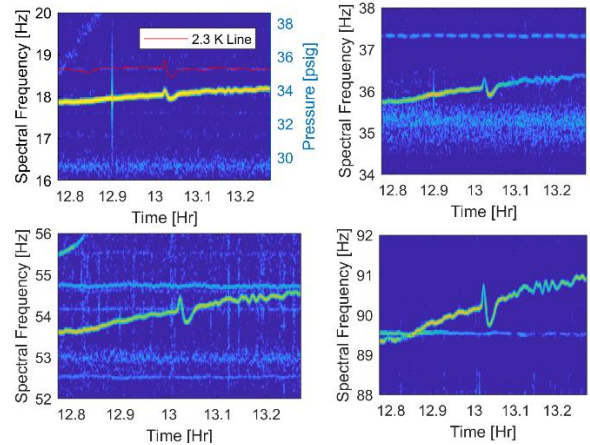

Figure 4: Spectrograms showing harmonics observed in cavity 1 in CM15. The fundamental mode occurs at 18 Hz.

For a pipe with one end open, the length is $L = n\lambda/4$. For n=1 the length is 3.125 m. The length of a pipe to create harmonics with these lengths is within the cryomodule's dimensions. Thus the harmonics observed in the frequency have a high probability of coming from the acoustic waves in helium.

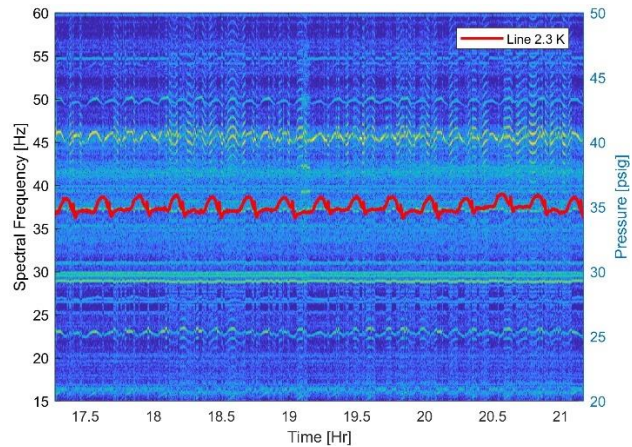

Figure 5: Spectrogram of cavity 5 in CM15 showing the harmonics with the fundamental frequency of 23 Hz and higher order frequency of 46 Hz.

## Effects of Liquid Helium Fluctuation

The smallest frequency excitations of the cavity occur due to the helium pressure variation in the cavity's liquid helium vessel, the correlation between these two parameters will be shown for different cryomodules. Other labs have observed this effect. A JT valve maintains the liquid level in the 2 K two-phase pipe and an external pump maintains the vapor pressure at 23 Torr to maintain it at 2 K. These two different mechanisms can give rise to the liquid helium pressure variation. The frequency pressure sensitivity (df/dp) measurements for the cavities yield a value of 75 +/- 5 Hz/torr, shown in Fig. 6. The effects of the helium pressure fluctuation on the cavity detuning are shown in Fig. 7. The variation is small, but it is enough to shift the cavity frequency.

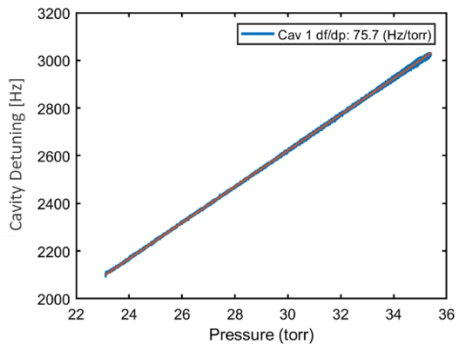

Figure 6: Cavity frequency sensitivity (df/dp).

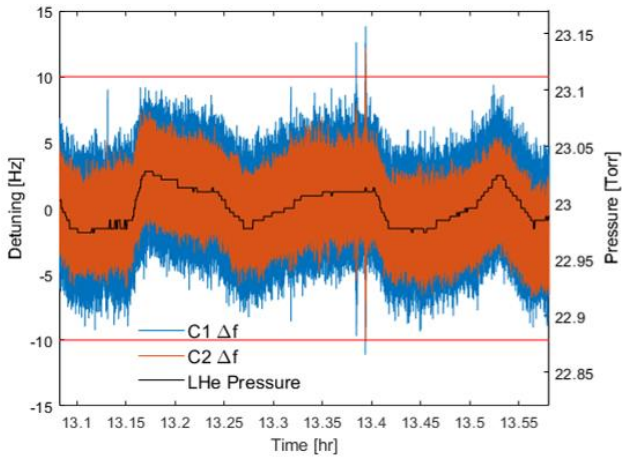

Figure 7: Cavity frequency detuning for cavities 1 and 2 on CM 17. The sampling rate of the helium pressure probe is 1 Hz.

Higher frequency excitations on the cavities can sometimes be correlated with the helium bath pressure. The correlation between sources > 280 Hz and the liquid helium pressure is shown Fig. 8 for CM 17. This data is the same as the one shown in Fig. 7. The peak detuning of all cavities was small, and only a few events were outside the 10 Hz limit (see Fig. 7). Cavity 7 had the strongest response to the detuning.

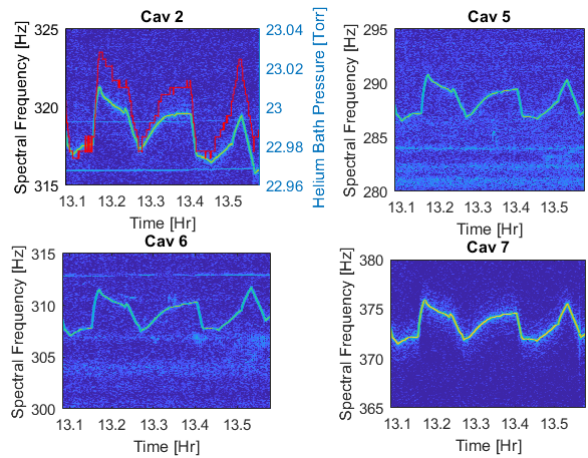

Figure 8: Cavity 7 spectrogram of helium bath pressure for CM 17 (same data as in Fig. 7). There is a correlation between a ~ 300 Hz line and the helium pressure.

## CONCLUSION

The LCLS-II cryomodules changed their design to reduce the effects of microphonics. These changes were also implemented for the cryomodules in LCLS-II HE. Despite these passive mitigation techniques, the cavities still experience frequency detuning caused by residual microphonics. The frequency detuning of 96 cavities and multiple cryogenic parameters were analyzed for 14 cryomodules. The results demonstrate that cryogenic parameters such as liquid helium level and supply helium pressure variation affect the frequency detuning.